# Interdisciplinarity in Socio-economics, mathematical analysis and predictability of complex systems

Didier Sornette

ETH Zurich
Department of Management, Technology and Economics
Zurich, Switzerland

*Correspondence*: dsornette@ethz.ch

**Abstract**: In this essay, I attempt to provide supporting evidence as well as some balance for the thesis on 'Transforming socio-economics with a new epistemology' presented by Hollingworth and Müller (2008). First, I review a personal highlight of my own scientific path that illustrates the power of interdisciplinarity as well as unity of the mathematical description of natural and social processes. I also argue against the claim that complex systems are in general 'not susceptible to mathematical analysis, but must be understood by letting them evolve—over time or with simulation analysis'. Moreover, I present evidence of the limits of the claim that scientists working within Science II do not make predictions about the future because it is too complex. I stress the potentials for a third 'Quantum Science' and its associated conceptual and philosophical revolutions, and finally point out some limits of the 'new' theory of networks.

# 1. Introduction

The essay of Rogers Hollingworth and Karl Müller (2008) on a new scientific framework (Science II) and on the key role of transfers-across-disciplines makes fascinating reading. As an active practitioner of several scientific fields (earthquake physics and geophysics, statistical physics, financial economics, and some incursions in biology and medicine), I witness everyday first-hand the power obtained by back-and-forth transfer of concepts, methods, and models occurring in interdisciplinary work and thus applaud the formalization and synthesis offered by Hollingworth and Müller (2008). The next section 2 presents a personal highlight of my own scientific path, that illustrates the power of interdisciplinarity as well s unity of the mathematical description of natural and social processes

But my goal is not just to flatter Rogers and Karl and praise their efforts. I wish here to suggest corrections and complements to the broad picture painted in Hollingworth and Müller (2008). I will discuss two major claims in some details. First, in section 3.1, I take issue with the claim that complex systems are in general 'not susceptible to mathematical analysis, but must be understood by letting them evolve—over time or with simulation analysis'. In section 3.2, I present evidence of the limits of the claim that scientists working within Science II do not make predictions about the future, because it is too complex. I conclude with section 4, in which I point out a possible missing link between Science I and Science II, namely 'Quantum Science', and the associated conceptual and philosophical revolution. I also tone down the optimism echoed by Hollingworth and Müller (2008) that the approaches in terms of complex networks will allow for a stronger transfer of theoretical models across widely disparate fields, in particular between the natural and social sciences.

# 2. A personal highlight illustrating the power of interdisciplinarity and unity of the mathematical description of natural and social processes

Let me illustrate with a personal experience how the power of interdisciplinarity can go beyond analogies to create genuinely new paths to discovery. In the example I wish to relate, the same fundamental concepts have been found to apply efficiently to model on the one hand the triggering processes between earthquakes leading to their complex space-time statistical organization (Helmstetter *et al.*, 2003; Ouillon and Sornette, 2005; Sornette and Ouillon, 2005) and on the other hand the social response to shocks in such examples as Internet downloads in response to information shocks (Johansen and Sornette, 2000), the dynamic of sales of book blockbusters (Sornette *et al.*, 2004; Deschatres and Sornette, 2005) and of viewers activity on YouTube.com (Crane and Sornette, 2007), the time response to social shocks (Roehner *et al.*, 2004), financial volatility shocks (Sornette *et al.*, 2003), and financial bubbles and their crashes (Johansen *et al.*, 1999, 2000; Sornette, 2003; Andersen and Sornette, 2005; Sornette and Zhou, 2006). The research process developed as follows.

First, the possibility that precursory seismic activity, known as foreshocks, could be intimately related to aftershocks has been entertained by several authors in the past decades but has not been clearly demonstrated by a combined derivation of the so-called direct Omori law for aftershocks and of the inverse Omori for foreshocks within a consistent model. In a nutshell, the Omori law for aftershocks describes the decay rate of seismicity after a large earthquake (called a mainshock), roughly going as the inverse of time since the mainshock. The inverse Omori law for foreshocks describes the statistically increasing rate of earthquakes going roughly as the inverse of the time till the next mainshock. The inverse Omori law has been demonstrated empirically only by stacking many earthquake sequences (see Helmstetter and Sornette, 2003 and references therein). We had been working for several years on the theoretical understanding of a statistical seismicity model, known at the ETAS model, a self-excited Hawkes conditional point process in mathematical parlance. This model or its siblings are now used as the standard benchmarks in statistical seismology and for evaluating other earthquake forecast models (Jordan, 2006; Schorlemmer *et al.*, 2007a, 2007b). We had the intuition that the inverse Omori law for foreshocks could be derived from the direct Omori law by viewing mainshocks as the 'aftershocks of foreshocks, conditional on the magnitude of mainshocks being larger than that of their progenitors'. But we could not find the mathematical trick to complete the theoretical derivation. In parallel, we were working on the statistical properties of financial returns and were starting a collaboration with J.-F. Muzy, one of the discoverers of a new stochastic random walk with exact multifractal properties, named the multifractal random walk (MRW; Bacry *et al.*, 2001; Muzy and Bacry, 2002), which seems to be a powerful model of financial time series. We then realized that similar questions could be asked on the precursory as well as posterior behavior of financial volatility around shocks. The analysis of the data showed clear Omori-like and inverse Omori-like behavior around both exogenous (Sept. 11, 2001, or the coup against Gorbachev in 1991) and endogenous shocks. It turned out that we were able to formulate the solution mathematically within the formalism of the MRW and we showed the deep link between the precursory increase and posterior behavior around financial shocks (Sornette *et al.*, 2003). In particular, we showed a clear quantitative relationship between the relaxation after an exogenously caused shock and the relaxation following a shock arising spontaneously (termed 'endogenous'). Then, inspired by the conceptual path used to solve the problem in the financial context, we were able to derive the solution in the context of the ETAS model, demonstrating mathematically the deep link between the inverse Omori law for foreshocks and the direct Omori law for aftershocks in the context of the ETAS model (Helmstetter *et al.*, 2003; Helmstetter and Sornette, 2003). The path was simpler and clearer for financial time series and their study clarified the methodology to be used for the more complicated specific point processed modeling earthquakes.

This remarkable back-and-forth thought process between two a priori very different fields will remain a personal highlight of my scientific life.

**3. On self-organizing processes and multi-level analysis**

The emphasis of Hollingworth and Müller (2008) on self-organizing processes and multi-level analysis to comprehend the nature of complex social systems is welcome as it indeed reflects an important strategy used by researchers. But more problematic is the endorsement of the claims, which are variations of a common theme, that

(1) 'increasingly analysts maintain that such systems are not susceptible to mathematical analysis, but must be understood by letting them evolve—over time or with simulation analysis',

(2) 'the emerging perspective, rapidly diffusing across academic disciplines, suggests that the world does not change in predictable way',

(3) 'hardly any scientist in these fields is able to make successful predictions about the future, as self-organizing processes are understood best by retrospective analysis'.

Hollingworth and Müller (2008) give thus resonance to a view upheld by various groups in different communities, which I find misguided and dangerous, while unfortunately widespread.

### 3.1 On models of complex systems

Let me first address claims (1) and (2), perhaps best personified by Stephan Wolfram and elaborated in his massive book entitled 'A New Kind of Science' (Wolfram, 2002). According to Wolfram, the most interesting problems presented by nature (biological, physical, societal) are likely to be formally undecidable or computationally irreducible, rendering proofs and predictions impossible. Take the example of the Earth's crust and the problem of earthquake prediction or the economies and financial markets of countries and the question of predicting their recessions and their financial crashes. Because these events depend on the delicate interactions of millions of parts, and seemingly insignificant accidents can sometimes have massive repercussions, it is argued that their inherent complexity makes such events utterly unknowable and unpredictable. To understand precisely what this means, let us refer to the mathematics of algorithmic complexity (Chaitin, 1987), which provides one of the formal approaches to the study of complex systems. Following a logical construction related to that underpinning Gödel's (1931) incompleteness theorem, most complex systems have been proved to be computationally irreducible, i.e. the only way to decide about their evolution is to actually let them evolve in time. The only way to find out what will happen is to actually let it happen. Accordingly, the future time evolution of most complex systems appears inherently unpredictable. Such statement plays a very important role in every discussion on how to define and measure complexity.

However, it turns out that this and other related theorems (see Chaitin, 1987 and Matthew Cook in Wolfram, 2002) are useless for most practical purposes and are in fact misleading for the development of scientific understanding. And the following explains why. In a now famous essay entitled 'More Is Different', Phil Anderson (1972), 1977 Nobel Prize winner and a founder of the Santa Fe Institute of complexity, described how features of organization arise as an 'emergent' property of systems, with completely new laws describing different levels of magnification. As a consequence, Physics for instance works and is not hampered by computational irreducibility. This is because physicists only ask for answers at some coarse-grained level (see Buchanan, 2005 for a pedagogical presentation of these ideas). In basically all sciences, one aims at predicting *coarse-grained* properties. Only by ignoring most of molecular detail, for example, did researchers ever develop the laws of thermodynamics, fluid dynamics and chemistry, providing remarkable tools for explaining and predicting new phenomena. From this perspective, one could say that the fundamental theorems of algorithm complexity are like pious acts of homage to our intellectual ancestors: they are solemnly taken out, exhibited, and solemnly put away as useless for most practical applications. The reason for the lack of practical value is the focus on too many details, forgetting that systems become coherent at some level of description. In the same vein, the butterfly effect, famously introduced by E. Lorenz (1963; 1972) to communicate the concept of sensitivity upon initial conditions in chaos, is actually not relevant to explain and predict the coherent meteorological structures at large scales. As a result of the spontaneous organization of coherent structures (Holmes *et al.*, 1998), there is actually predictability in meteorology and climate as well as in many other systems, at time scales of months to years, in apparent contradiction with the superficial insight provided by the butterfly effect.

These points were made beautifully clear by Israeli and Goldenfeld (2004; 2006), in their study of cellular automata, the very mathematical models that has led Wolfram to make his grand claims that science should stop trying to make predictions and scientists should only run cellular automata on their computers to reproduce, but not explain, the complexity of the world. Cellular automata are systems defined in discrete Manhattan-like meshed spaces and evolve in discrete time steps, with discrete-valued variables interacting according to simple rules. These remarkable simple systems have been shown to be able to reproduce many of the behavior of complex systems. In particular, it is known that most of them are 'universal Turing computational machines', i.e. they are capable of emulating any physical machine. Because they can emulate any other computing device, they are therefore undecidable and unpredictable. But which of the systems are capable of universal computation is not generally known. In this respect, one results stands out, for our purpose. Matthew Cook, whose theorem is reported in (Wolfram, 2002), showed that one simple cellular automaton, known as 'rule 110' in Wolfram's nomenclature of one-dimensional cellular automata with nearest-neighbor interaction rules, is such a universal Turing machine.

Now, Navot Israeli and Nigel Goldenfeld applied a technique called 'renormalization group' (Wilson, 1999) to search for what could be the new laws, if any, that describe the coarse-grained average evolution of such cellular automata. Technically, the new laws are determined by a self-consistency condition that (i) coarse-graining the initial conditions and applying the new laws should provide the same final description as (ii) letting evolve the system according to the true microscopic laws and then coarse-graining the resulting pattern. By coarse-graining, one focuses only on the most relevant details of the pattern-forming processes. Israeli and Goldenfeld established that computationally irreducible cellular automata become predictable and even computationally reducible at a coarse-grained level of description. The resulting coarse-grained cellular automata that they constructed by coarse-graining different cellular automata were found to emulate the large-scale behavior of the original systems without accounting for small-scale details. In particular, rule 110 was found to become a much simpler predictable system, upon coarse-graining. By developing exact coarse-grained procedures on computationally irreducible cellular automata, Israeli and Goldenfeld have demonstrated that a scientific predictive theory may simply depend on finding the right level for describing the system. For physicists, this is not a surprise: by asking only for approximate answers, Physics is not hampered by computational irreducibility, and I believe that this statement holds for all natural and social sciences with empirical foundations.

### 3.2 On predictability of the future in complex systems

Let me now turn to the third claim cited in the above introduction of section 3 that 'hardly any scientist in these fields is able to make successful predictions about the future', and more generally that predicting the future from the past is inherently impossible from most complex systems. This view has recently been defended persuasively in concrete prediction applications, such as in the socially important issue of earthquake prediction (see e.g. the contributions in Nature debate [1999]). In addition to the persistent failure in reaching a reliable earthquake predictive scheme up to the present day, this view is rooted theoretically in the analogy between earthquakes and self-organized criticality (Bak, 1996). Within this 'fractal' framework, there is no characteristic scale and the power law distribution of earthquake sizes suggests that the large earthquakes are nothing but small earthquakes that did not stop. Large earthquakes are thus unpredictable because their nucleation appears to be not different from that of the multitude of small earthquakes.

Does this really hold for all features of complex systems? Take our personal life. We are not really interested in knowing in advance at what time we will go to a given store or drive in a highway. We are much more interested in forecasting the major bifurcations ahead of us, involving the few important things, like health, love and work that count for our happiness. Similarly, predicting the detailed evolution of complex systems has no real value and the fact that we are taught that it is out of reach from a fundamental point of view does not exclude the more interesting possibility to predict the phases of evolutions of complex systems that really count (Sornette, 1999).

It turns out that most complex systems around us do exhibit rare and sudden transitions, that occur over time intervals that are short compared with the characteristic time scales of their prior or posterior evolution. Such extreme events express more than anything else the underlying 'forces' usually hidden by almost perfect balance and thus provide the potential for a better scientific understanding of complex systems. By focusing on these characteristic events, and in the spirit of the coarse-graining metaphor of the cellular automata discussed in section 3.1, a small but growing number of scientists are re-considering the claims of unpredictability. After the wave of complete pessimism on earthquake prediction in the West of the 1980s and 1990s, international earthquake prediction experiments such as the recently formed Collaboratory for the Study of Earthquake Predictability (CSEP; Jordan, 2006) and the Working Group on Regional Earthquake Likelihood Models (RELM; Schorlemmer *et al*., 2007a, 2007b) aim to investigate scientific hypotheses about seismicity in a systematic, rigorous and truly prospective manner by evaluating the forecasts of models against observed earthquake parameters (time, location, magnitude, focal mechanism, etc) that are taken from earthquake catalogs.

Recent developments suggest that non-traditional approaches, based on the concepts and methods of statistical and nonlinear physics could provide a middle way to direct the numerical resolution of more realistic models and the identification of relevant signatures of impending catastrophes, and in particular of social crises. Enriching the concept of self-organizing criticality, the predictability of crises would then rely on the fact that they are fundamentally outliers (Johansen and Sornette, 2001), e.g. financial crashes are not scaled-up versions of small losses, but the result of specific collective amplifying mechanisms (see chapter 3 in Sornette, 2003, where this concept is documented empirically and discussed in the context of coherent structures in hydrodynamic turbulence and of financial market crashes). To address this challenge, the available theoretical tools comprise in particular bifurcation and catastrophe theories, dynamical critical phenomena and the renormalization group, nonlinear dynamical systems, and the theory of partially (spontaneously or not) broken symmetries. Some encouraging results have been gathered on concrete problems (see the reviews Sornette, 2005, 2008 and references therein), such as the prediction of the failure of complex engineering structures (a challenge generally thought unreachable by most material scientists), the detection of precursors to stock market crashes with real advance published predictions (another unattainable challenge generally according to most financial economists) and the prediction of human parturition and epileptic seizures, to cite some subjects I have been involved with, with exciting potential for a variety of other fields.

Other pioneers in different disciplines are slowly coming up to grip with the potential for a degree of predictability of extreme events in many complex systems (see, for instance, the chapters in Albeverio *et al*., 2005). Let us also mention Jim Crutchfield who proposes that connections between the past and future could be predicted for virtually any system with a 'computational mechanics' approach based on sorting various histories of a system into classes, so that the same outcome applies for all histories in each class (Ay and Crutchfield, 2005; Crutchfield and Görnerup, 2006). Again, many details of the underlying system may be inconsequential, so that an approximate description much like Isreali and Goldenfeld's coarse-grained cellular automata models can be organized and used to make predictions.

Agent-based models developed to mimic financial markets have been found to exhibit a special kind of predictability. While being unpredictable most of the time, these systems show transient dynamical pockets of predictability in which agents collectively take predetermined courses of action, decoupled from past history. Using so-called minority and majority games as well as real financial time series, a surprisingly large frequency of these pockets of predictability have been found, implying a collective organization of agents and of their strategies which 'condense' into transitional herding regimes (Lamper *et al.*, 2002; Andersen and Sornette, 2005). Again, grand claims of intrinsic lack of predictability seem to me like throwing out the baby with the bath water, forgetting that the heterogeneous nature in space and time of the self-organization of complex systems does not exclude partial predictability at some coarse-grained level.

Let me end this discussion by extrapolating and forecasting that a larger multidisciplinary integration of the physical and social sciences together with artificial intelligence and soft-computational techniques, fed by analogies and fertilization across the natural and social sciences, will provide a better understanding of the limits of predictability of crises.

**4. Concluding remarks on Quantum Decision Theory and the Theory of Networks**
I would like to conclude with two remarks.

Hollingworth and Müller (2008) contrast the 'old' Descartes–Newton Science I with the new Science II framework which emphasizes concepts such as complex adaptive systems, self-organization and multi-scale patterns, scale invariance, networks and other buzzwords. I was surprised not to see discussed another Science, the 'Quantum Science' emerging from the scientific and philosophical revolution triggered by the understanding that Nature works through the agency of fundamentally quantum mechanical laws, which have very little to do with the macroscopic laws apparent directly to our five perception senses. In my view, for a variety of disciplines, but perhaps not yet for the social sciences, quantum mechanics has had more impact than Science II. At the ontological level, Quantum Science has had a tremendous influence in all fields, by providing a fundamental probabilistic framework, rooted in the Heisenberg uncertainty principle, the intrinsic non-separability theorem and the existence of intrinsic sources of noise and energy in the fluctuations of the 'void' (showing that the void does not exist ontologically). This has attacked, much more deeply than e.g. the theory of chaos ever has, the misconception that future scenarios are deterministic and fully predictable.

In this concluding section, I would like to suggest that Quantum Science may enjoy a growing impact in the social sciences, via the channel of decision making operating in humans by emphasizing the importance of taking into account the superposition of composite prospects, whose aggregated behavior form the structures, such as society and economies, that scholars strive to understand. In a preliminary essay, Slava Yukalov and I have introduced a 'quantum decision theory' (QDT) of decision making based on the mathematical theory of separable Hilbert spaces on the continuous field of complex numbers (Yukalov and Sornette, 2008), the same mathematical structure on which quantum mechanics is based. This mathematical formulation captures the effect of superposition of composite prospects, including many incorporated intentions, which allows us to describe a variety of interesting fallacies and anomalies that have been reported to characterize the decision making processes of real human beings.

My second remark concerns the claim that 'complex networks allow for a transfer of theoretical models across widely disparate fields'. I am afraid that the optimism that the theory of complex networks will play such a special role is nothing but another hype, somewhat in the lineage of those in the last decades that involved buzzwords such as fractals, chaos, self-organized criticality. They all had their period of fame and excesses, followed by maturation towards a more reasonable balanced position within the grand edifice of science. With Max Werner, we have recently commented on the limits of applying network theory in the field of earthquake modeling and predictability (Sornette and Werner, 2008) and I believe much the same criticisms would apply to the use of network theory to the social sciences. With Yannick Malevergne and Alex Saichev, we have developed in theoretical synthesis (Malevergne *et al.*, 2008) showing in particular that the mechanism of 'preferential attachment', at the basis of the understanding of scale-free networks found in social networks, the world-wide web, or networks of proteins reacting with each other in the cell, is nothing but a rediscovery and rephrasing in a slightly different language of the famous model of incoming and growing firms developed by Simon in 1955, based on Gibrat principle of proportional growth (Gibrat, 1931). The 'new' science of networks has thus deep roots in economics! Viewing the rather unsophisticated level of many discussions on power laws and other statistical regularities reported in the 'new' science of networks, while not disputing the existence of significant progress in network theory, I wonder whether this 'new' science would not profit from a better reading of the best works in economics of the twentieth century. Ending on a more positive note, this illustrates my fervent faith in the power of interdisciplinarity, practiced with the rigor and diligence necessary to ensure depth and fecundity.